\documentclass{article}
\usepackage{newsprocl}
\begin{document}
\title{Elements of information-theoretic derivation of the formalism of quantum theory.}

\author{A. GRINBAUM}
\address{CREA, Ecole Polytechnique,\\
  1 rue Descartes 75005 Paris, France\\
  E-mail: grinbaum@poly.polytechnique.fr}

\maketitle
\begin{abstract}
Information-theoretic derivations of the formalism of quantum
theory have recently attracted much attention. We analyze the
axioms underlying a few such derivations and propose a conceptual
framework in which, by combining several approaches, one can
retrieve more of the conventional quantum formalism.
\end{abstract}

\section{Strategy}\label{sect1}

Initially formulated by John Wheeler\cite{WheIBM,WheFeynman},
the program of deriving quantum formalism from
information-theore\-tic principles has been receiving lately much
attention. Thus, Jozsa\cite{Jozsa} promotes a viewpoint which
"attempts to place a notion of information at a primary
fundamental level in the formulation of quantum physics". Fuchs
\cite{FuchsLittleMore} presents his program as follows: "The task
is not to make sense of the quantum axioms by heaping more
structure, more definitions... on top of them, but to throw them
away wholesale and start afresh. We should be relentless in asking
ourselves: From what deep \emph{physical} principles might we
\emph{derive} this exquisite mathematical structure?.. I myself
see no alternative but to contemplate deep and hard the tasks, the
techniques, and the implications of quantum information theory."

In a similar fashion, Rovelli\cite{RovRQM} distinguishes a
philosophical problem of interpretation from a mathematical
problem of derivation of quantum mechanical formalism from the
first principles. He writes, "... quantum mechanics will cease to
look puzzling only when we will be able to \textit{derive} the
formalism of the theory from a set of simple physical assertions
('postulates', 'principles') about the world. Therefore, we should
not try to append a reasonable interpretation to the quantum
mechanics \textit{formalism}, but rather to \textit{derive} the
formalism from a set of experimentally motivated postulates".
Rovelli refers to his own work as a point of view and not as
interpretation: "From the point of view discussed here, Bohr's
interpretation, consistent histories interpretations, as well as
many worlds interpretation, are all correct". Rovelli's
\emph{point of view}, i.e.  informational treatment of quantum
mechanics, thus serves a formal criterion or a filter that permeates
certain interpretations and not others. In other words, treatment
of quantum mechanics on information-theoretic grounds entails that
some interpretations of quantum theory will be with certainty
inapplicable but a number of other interpretations will all remain
possible. Such a result can be naturally expected from any novel
formal development of quantum theory that remains in the area of
science as opposed to philosophy.

\section{Axioms}

\subsection{The choice of axioms}

Any formal derivation of quantum mechanics, in particular those
using Bayes\-ian methods\cite{BCFFS,SBC,CFS,FuchsLittleMore} and
quite promising for someone who believes in informat\-ion-theor\-etic
foundations of physics, requires a definite conceptual background
on which such a derivation will further operate. As it is often
the case, to give a rigorous axiomatic system that could provide
the necessary background, is a difficult task. Below we analyze
some three proposed solutions, by Rovelli\cite{RovRQM}, by Fuchs
\cite{FuchsLittleMore,FuchsInLight} and by Brukner and Zeilinger
\cite{BZ}.

Elementary act of measurement is understood by Rovelli as yes-no
question. Brukner and Zeilinger use the term "proposition" which
generalizes the notion of binary question. Still, if one looks
into where from the term "proposition" appears, one finds in
\cite{ZeilFound} two formulations of Zeilinger's fundamental
principle for quantum mechanics:
\begin{description}
    \item[FP1] An elementary system represents the truth value of
    one proposition.
    \item[FP2] An elementary system carries one bit of
    information.
\end{description}
It seems that Zeilinger's choice of these two principles strongly
suggests that the following phrase in BZ reflects the view of the
authors on the fundamental issue and thus puts them very close to
Rovelli's position: "Yes-no alternatives are representatives of
basic fundamental units of all systems."

Fuchs starts directly with the Hilbert space and the full
structure of quantum mechanics. He desribes measurements not by
projectors but by positive operator-valued measures. This allows
one to think that he will not agree with a definition of primitive
measurements as consisting of exclusive yes-no alternatives, where
the word "exclusive" leads to mathematically representing yes-no
questions as orthogonal projectors. Still, Fuchs mentions some of
the basic assumptions that he makes in his derivation.

Rovelli and BZ each then pose two axioms.
\begin{description}
    \item[Axiom 1] :\par\begin{itemize}
    \item Rovelli: "There is a maximum amount of relevant information that can be
extracted from a system."
    \item Fuchs: Doesn't follow the axiomatic approach; states that "There is maximal information about a system."
    \item Brukner and Zeilinger: "The information content of a quantum system is finite."
\end{itemize}
    \item[Axiom 2] :\par\begin{itemize}
    \item Rovelli: "It is always possible to acquire new information about a
    system."
    \item Fuchs: Doesn't follow the axiomatic approach; states that 
"There will always be questions that we can ask of a system for
which we cannot predict the outcomes."
    \item Brukner and Zeilinger: Introduce the notion of total information content of the
    system; state that there exist mutually complementary propositions; state that
    total information content of the system is invariant under a change of the set of
mutually complementary propositions.
\end{itemize}
\end{description}

In spite of a quite striking analogy between the axioms chosen by
different authors, as for the following derivation of quantum
mechanics, they do not proceed in the same manner. We shall now
have a closer look at the axioms and derivation techniques.

\subsection{Discussion of the axioms}

Axiom 1 marks a crucial point of departure from classical physics.
Newtonian physics employs mathematics of continuum to represent
the world and, therefore, any calculation of complete information
about, say, a particle position would require an infinitely long
computation. This fact has profoundly disturbed many physicists,
with most prominently Feynman saying, "It always bothers me that,
according to the laws as we understand then today, it takes a
computing machine an infinite number of logical operations to
figure out what goes on in no matter how tiny a region of space
and no matter how tiny a region of time,... why should it take an
infinite amount of logic to figure out what one tiny piece of
space-time is going to do?" Axiom 1 also goes in line with
Wheeler's "no continuum" principle\cite{WheFeynman}.

While it seems intuitively plausible to accept Axiom 1, its
interpretation is not straightforward. Each author imposes his own
interpretation by choosing a suitable translation into his
language; for Rovelli it means that there exist complete sets of
yes-no questions that could provide one, abstractly, with complete
information about a system; for Brukner and Zeilinger it means
that everything that is there to a system is represented by a
complete set of mutually complementary propositions.

However, it appears that Axiom 1 can raise yet a different issue.
Our intuition is that essential finiteness applies, not to the
system to which we address yes-no questions but to the system plus
the observer who asks these questions. Formal development of this
idea will appear in section \ref{sect3}. Philosophical argument
goes as follows: it is not true, that in order to know a Newtonian
coordinate we need the knowledge of infinitely many decimal
digits. The latter should not make us worry, for we are endowed
with an ability to create a special code (a new concept), which
will substitute in the thinking the undesired infinity. The same
works for computation: Feynman's argument from the infinity of
logical operations must include the possibility of "hiding"
Newtonian infinities under the "and so forth" concept, for which
one may specify operational rules. Consequently, the requirement
of finiteness applies to the observer-observed system. In a
similar manner, in BZ terms, essential finiteness applies to
system plus the one who chooses the propositions to be tested on
the system, i.e. the observer. We do not have the intuition that
"one cannot know the infinity" but, rather, that one cannot have
infinite knowledge.

How is this understanding of the finiteness axiom related to the
one adopt\-ed by Rovelli and other authors? With the assumption of
universality of quantum theory, one can deduce the old point of
view. Indeed, universality allows us to treat the border line
between theory and meta-theory in quantum mechanics as flexible
\cite{DallaChiara}. Any given observer (meta-theoretical entity)
can be included in the theory proper by taking the point of view
of an observer external to the one in question. If the amount of
information remains finite in spite of the arbitrary choice of the
frontier between the system and the observer, we can eliminate
from consideration any previously given observer at the price of
redefining the question-asking party. Imagine, for example, a
computer solving in Maple software some field theory
renormalization problem. Renorm\-aliz\-at\-ion is about removing
infinities, so if a computer were let to solve this task without
conceptualizing infinities by means of a previously learned
renormalization technique, it would have never arrived at any
result. Abstract amount of information in the system is infinite
in this example, but the amount of relevant information is finite.
What is relevant, decides the observer who translates relevancy
into concepts that he employs for the computation on a system, or
equivalently into a specific manner to ask some yes-no questions
and not others (see section \ref{sect44}).

Going to the extreme of definitions, what is information and what
is not decides the observer, and it is because of this that the
amount of information is finite. Had we had the liberty to call
information anything we want, there would be no intuitively clear
argument showing why this "anything" must be finite. Finiteness
thus has to include the observer, and thanks to the universality
of quantum mechanics can be in the limit reduced to finiteness in
the sense of Rovelli.

Axiom 2 beautifully corresponds to Wheeler's dictum (adopted after
Philip Anderson) "More is different". If one wants to get more
information, this will be different information; or one can always
get more information, and it will be different information. Though
in the original "More is different" was used in the context of
complexity theory, it can as well, as a basic principle, apply to
information-based quantum theory. At it will be seen below, for
Rovelli Axiom 2 allows to introduce probabilities and to deduce an
analogue of the Born rule; for BZ Axiom 2 leads to imposing a
certain structure on the information space. This axiom is
responsible for the departure from classicality, which is not yet
fully accommodated by Axiom 1 alone.

\subsection{Possible development}

Fuchs uses as \textit{a priori} given the structure of Hilbert
space; his task is to deduce some of the operational structure of
quantum mechanics, namely, density matrices. Rovelli, on the
contrary, is interested in deducing quantum mechanics from the
axioms and does not show a way to deduce most of further
structure, to start with the superposition principle (apart from
introducing it as an axiom). Fuchs uses a decision-theoretic
(Bayesian) approach to derive the superposition principle. He
refers to Rovelli's paper in his own, and one is left free to
suggest that many of his axiomatic assumptions, on which he
doesn't clearly comment, might be similar to Rovelli's ones, apart
from the key issue of how to define measurement. Indeed, Fuchs
insists on the fundamental character of positive operator-valued
measures. This may not seem intuitively evident. But because Fuchs
leaves the axiomatic foundations of the Bayesian approach open,
even if we dismiss the necessity to define measurement as POVM,
there still remains an opportunity to introduce the latter in the
theory. POVM have a natural description as conventional von
Neumann measurements on an ancilla system\cite{Peres}, and thus
to Rovelli's axiomatic derivation of the Hilbert space structure
one may try to add an account of inevitability of ancilla systems
and naturally obtain from this the needed POVM description. This
will be attempted in section \ref{sect3}.

Brukner and Zeilinger proceed differently. If information is
primary, they argue, then any formalism should be a formalism
dealing with information and not with some other notions.
Therefore BZ construct an \textit{information space }where they
apply the axioms and use the formalism to deduce testable
predictions. BZ do not refer to physical space or to Hilbert space
in their construction. Thus they do not have access to algebra
allowing a reconstruction of the state space out of the operator
Hilbert space. Therefore, because of this change of scenery, they
are bound to postulate more properties of mutually complementary
propositions than Rovelli or Fuchs. Namely, they postulate the
homogeneity of parameter space. BZ's self-imposed terminological
limitation to abstract information space does not seem viable for
philosophical, i.e. extra-scientific, reasons: it renders the
formalism less transparent in use, while introduction of
supplementary axioms does not make it conceptually clearer than
traditional formalisms.

To continue, the question is how to extract a useful approach from
the juxtaposition of Rovelli's and Fuchs's proposals. Rovelli, as
said before, shows a way to construct the Hilbert space structure
from two axioms. Unlike Brukner and Zeilinger, we do not call this
Hilbert space information space but simply physical space, for
there is no other space in the whole construction that would be
the physical space. BZ's information space is what the physical
space \emph{is}\footnote{Our use of the verb \textit{to be} does
not imply that we hold any form of realism. We merely refer here
to common usage of the term \textit{physical space}.}. Next, with
Hilbert space in hand, we use Fuchs's derivation based on
Gleason's theorem to deduce density matrices and their properties.
This requires essentially one more step: we need to introduce POVM
as measurements, as discussed above. Once we've introduced ancilla
systems, we can operationally redefine measurement as described by
POVM.

To move further in combining Rovelli's and Fuchs's proposals,
after the axiomatic stage, we either need a sort of
decision-theoretic approach to derive the formal consequences of
the necessary intersubjective accord of measurement results (for
Fuchs, for example, via a version of the de Finetti theorem), or
we need to use an algebraic approach so that the constructed
Hilbert space be treated as space of operators corresponding to
observables. This latter option will be investigated below.

\section{Reconstruction of the quantum formalism}\label{sect3}

\subsection{Key metaphor}

We are guided by the computer metaphor. Indeed, the strategic task
is to give a reformulation of quantum theory in
information-theoretic terms. A theory that operates with the
notion of information can be compared to software as opposed to a
theory that operates with the notion of energy which can be
compared to hardware. Ideally one would wish to see all "hardware"
or energetic language disappear from the formulation of the
theory, so that only "software" or informational language remain.

\subsection{I-observer and P-observer}

We are usually interested in information about (knowledge of) the
chosen system and we disregard particular ways in which we have
obtained this information. All that counts is knowledge that can
be useful in future or, in other words, relevant knowledge or
relevant information. This is why one usually does not pay
attention to the very process of interaction between the system
being measured and the measuring system, and one treats measuring
system as a meta-theoretic, i.e. non-physical, apparatus. To give
an example, for some experiment a physicist may need to know the
proton mass but he will not at all be interested in how this
quantity was measured, unless he is a narrow specialist whose
interest is in measuring particle masses. Particular ways to gain
knowledge are irrelevant, while knowledge itself is highly
relevant and useful. Some of the experiments where one is
interested in the measurement as a physical process are discussed
in\cite{Mensky}. From now on we assume that measurement details
are irrelevant, perhaps at the price of redefining what is
measurement.

In a practical setting, though, information is \textit{always}
physical\cite{Landauer}. This is to say that there always is some
physical support of information, some hardware. The necessity of
the physical support requires that we proceed in the following
manner: first, treat the measurement interaction as physical;
then, disregard the fact that it was physical and reformulate the
theory in terms of measurement results only.

To start, make a distinction between two parts of the world:
quantum system $S$, which is the system of interest, and the
observer. The observer, in the spirit of the software-hardware
metaphor, consists of an informational agent ("I-observer") and of
the physical realization of the observer ("P-observer"). There is
no I-observer without P-observer. Reciprocally, there is no sense
in calling P-observer an observer unless there is I-observer
(otherwise P-observer is just a physical system as any). Hence,
the two components of the "larger observer" are not in any way
separate or orthogonal to each other; on the contrary, these are
merely two viewpoints, and the difference is but descriptive.

\subsection{Hilbert space}

P-observer interacts with the quantum system and thus provides for
the physical basis of measurement. I-observer is only interested
in the measurement result, i.e. information per se, and he gets
information by reading it from P-observer. The act of reading or
getting information is here a common linguistic expression and not
a physical process since I-observer and P-observer are not
physically distinct. In fact, the concept of "being physical" only
applies to P-observer, and by definition the physical content of
the "larger observer" is all contained in P-observer. I-observer
as informational agent is meta-theoretic, and hence the fact that
its interaction with P-observer, or the act of "reading
information", is unphysical. To give a mathematical meaning to
this act, we assume that getting information is described as
yes-no questions asked by I-observer to P-observer.

To follow Rovelli's construction\cite{RovRQM}, the set of
questions will be denoted $W(P)=\{Q_i, i\in I\}$. According to
Axiom 1, there is a finite number $N$ that characterizes
P-observer's informational capacity. The number of questions in
$I$, though, can be much larger than $N$, as some of these
questions are not independent. In particular, they may be related
by implication ($Q_1\Rightarrow Q_2$), union ($Q_3=Q_1\vee Q_2$),
and intersection ($Q_3=Q_1\wedge Q_2$). One can define an always
false ($Q_0$) and an always true question ($Q_\infty$), negation
of a question ($\neg Q$), and a notion of orthogonality as
follows: if $Q_1 \Rightarrow \neg Q_2$, then $Q_1$ and $Q_2$ are
orthogonal ($Q_1 \bot Q_2$). Equipped with these structures, and
under the non-trivial assumption that union and intersection are
defined for every pair of questions, $W(P)$ is an orthomodular
lattice.

One needs to make a few more steps to obtain the Hilbert space
structure. As follows from Axiom 1, one can select in $W(P)$ a set
$c$ of $N$ questions that are independent from each other. In the
general case, there exist many such sets $c$, $d$, etc. If
I-observer asks the $N$ questions in the family $c$ then the
obtained answers form a string $$s_c=[e_1,\ldots,e_N]_c.$$ This
string represents the "raw" information that I-observer got from
P-\-observ\-er as a result of asking the questions in $c$. Note that
this is not yet information about the quantum system $S$ that the
I-observer ultimately wants to have, but only a process due to
functional separation within the "larger observer".

The string $s_c$ can take $2^N$ values and, since these outcomes
are by construction mutually exclusive, we can define new
questions $Q_c^{(1)}\ldots Q_c^{(2^N)}$ such that the yes answer
to $Q_c^{(i)}$ corresponds to the string of answers $s_c^{(i)}$:
$$Q_c^{(i)}=\neg Q_1 \wedge \neg Q_2 \wedge \ldots \wedge\neg
Q_{N-i+1}\wedge Q_{N-i+2}\wedge\ldots\wedge Q_N.$$ To these
questions we refer as to "complete questions". By taking all
possible unions of sets of complete questions $Q_c^{(i)}$ of the
same family $c$ one constructs a Boolean algebra that has
$Q_c^{(i)}$ as atoms.

Alternatively, one can consider a different family $d$ of N
independent yes-no questions and obtain another Boolean algebra
with different complete questions as atoms. It follows, then, from
Axiom 1 that the set of questions $W(P)$ that can be asked to
P-observer is algebraically an orthomodular lattice containing
subsets that form Boolean algebras. This is precisely the
algebraic structure formed by the family of linear subsets of
Hilbert space.

It is interesting to note that in approaches that start with an
abstract $C^*$-algebra of operators one needs to use the
Gelfand-Naimark-Segal construction to obtain a representation of
this algebra as algebra of operators on a Hilbert space. In the
present approach, information-theoretic axioms are evoked to
obtain a similar result, namely, to show that operators form a
Hilbert space.

\subsection{Born rule}

From the second Rovelli's axiom it follows immediately that there
are questions such as answers to these questions are not
determined by $s_c$. Define, in general, as $p(Q,Q_c^{(i)})$ the
probability that a yes answer to $Q$ will follow from the string
$s_c^{(i)}$. Given two complete strings of answers $s_c$ and
$s_b$, we can then consider the probabilities
$$p^{ij}=p(Q_b^{(i)},Q_c^{(j)}).\footnote{This introduction of
probabilities does not yet commit one to any particular view on
what probabilities \textit{are}. Personally, the author believes
in the trascendental deduction of the structure of probabilities
\cite{Petitot,Bitbol} and in the subjective attribution of numeric
values to probabilities.}$$ From the way it is defined, the
$2^N\times 2^N$ matrix $p^{ij}$ cannot be completely arbitrary.
First, we must have $$0\leq p^{ij}\leq 1.$$ Then, if information
$s_c^{(j)}$ is available about the system, one and only one of the
outcomes $s_b^{(i)}$ may result. Therefore $$\sum _i p^{ij}=1.$$
If we assume that $p(Q_b^{(i)},Q_c^{(j)})=p(Q_c^{(j)},Q_b^{(i)})$
then we also get $$\sum _j p^{ij}=1.$$

If pursued further in an attempt to deduce probability amplitudes,
this derivation, however, encounters some difficulties. To get the
result, Rovelli postulates explicitly the superposition principle.
We, too, introduce a new assumption to obtain more of the
structure of quantum theory. Namely, we postulate
non-contextuality and use Gleason's theorem to deduce density
matrices. It remains an open question if non-contextuality as an
intuitively made assumption is welcome or must be rejected as too
strong\cite{Saunders}. In mathematical terms, it states that
probabilities can be defined for a projector independently of the
family of projectors of which it is a member, or that in
$p(Q_b^{(i)},Q_c^{(j)})$ with fixed $Q_b^{(i)}$ probability will
be the same had the fixed question belonged not to the family $b$
but to some other family $d$. One can then prove a theorem due to
Gleason\cite{Gleason}:

\textbf{Theorem (Gleason)} \textit{Let $f$ be any function from
1-dimensional projections on a Hilbert space of dimension $d>2$ to
the unit interval, such that for each resolution of the identity
$\{ \Pi_k\}, k=1\ldots d, \sum _{k=1}^d \Pi_k = I, \sum _{k=1}^d
f(\Pi_k) = 1.$ Then there exists a unique density matrix $\rho$
such that $f(\Pi_k)=Tr(\rho \Pi_k).$}

\subsection{Unitary dynamics}\label{sect44}

One last step before we move to quantum theory of the system $S$
is to obtain unitary dynamics. Following Rovelli, any question can
be labelled by the time variable $t$ indicating the time at which
it is asked. Denote as $t\rightarrow Q(t)$ the one-parameter
family of questions defined by the same procedure performed at
different times. Assume that time evolution is a symmetry in the
theory.
 In the context of our approach the latter
word "theory" includes theory of P-observer \textit{and} of the
quantum system $S$. Then recall that the set $W(P)$ has the
structure of a set of linear subspaces in the Hilbert space, and
the set of all questions at time $t_2$ to the P-observer part of
the physically interacting conjunction of two systems, must be
isomorphic to the set of all questions at time $t_1$. Therefore,
the corresponding family of linear subspaces must have the same
structure; it follows that there must be a unitary transformation
$U(t_2-t_1)$ such that $$Q(t_2)=U(t_2-t_1)Q(t_1)U^{-1}(t_2-t_1).$$
It is straightforward to see that these unitary matrices form an
abelian group and $U(t_2-t_1)=\exp {[-\mathrm{i}(t_2-t_1)H]}$,
where $H$ is a self-adjoint operator in the Hilbert space, the
Hamiltonian.

In a practical setting, it is from the past or the future of a
given experiment, in particular from the intentions of the
experimenter, that one can learn which information about the
experiment is relevant and which is not. What is relevant can
either be encoded in the preparation of the experiment or selected
by the experimenter \emph{a posteriori}. In all cases, the notion
of relevance does not enter into the formalism which solely
describes the measurement within the context of the experiment.
All that is "allowed to be known" inside the formal framework is
that there (a) \textit{is} (b)\textit{ some} relevant information.
What is relevant is reflected in the choice of questions that are
asked by I-observer.

Interaction between P-observer and the quantum system should be
viewed as physical interaction between just any two physical
systems. Still, because I-observer then reads information from
P-observer and because we aren't interested in the posteriority of
relations between P-observer and the quantum system, we can treat
P-observer as an ancillary system in course of its interaction
with $S$. Such an ancillary system would have interacted with $S$
and then would be subject to a standard measurement described
mathematically on its Hilbert space via a set of orthogonal yes-no
projection operators.

So far, for P-observer we have the Hilbert space and the standard
Born rule. The fact that P-observer is treated as ancillary system
allows to transfer some of this structure on the quantum system
$S$. A new non-trivial assumption has to be made, that the time
dynamics that has previously arisen in the context of P-observer
alone, also applies to the I-observer and to $S$. In other words,
there is only one time in the system. Time of I-observer is the
one in which one can grasp the meaning of the words "past" and
"future" as used above in relation with the experimental setting
and the notion of relevant information: it is in this time that
there is a "before the experiment" and an "after the experiment".
Times of physical systems, such as $S$ or P-observer, are times in
which their dynamics takes place.

Now, both the physical interaction of P-observer with $S$ and the
process of asking questions by I-observer to P-observer take place
in one and the same time. Since (a) until I-observer asks the
question that he chooses to ask, sets of questions at different
times are isomorphic and evolution is unitary, and (b) time at
which I-observer asks the question only depends on I-observer and
considerations of relevance that must not enter into the
formalism, then one concludes that the interaction between the
quantum system and P-observer must respect the unitary character
all until the decoupling of the ancilla. Now write,
$$\rho_{SP}\rightarrow U\rho_{SP}U^\dag.$$ After asking a question
corresponding to a projector $\Pi _b$, probability of the yes
answer will be given by $$P(b)=Tr\left(U(\rho _S\otimes\rho
_P)U^\dag(I\otimes\Pi _b)\right).$$ Because the systems decouple,
trace can be decomposed into $$P(b)=Tr _S (\rho_S E_b),$$ where
all presence of the ancilla is hidden in the operator $$E_b=Tr
_P\left( (I\otimes\rho _P)U(I\otimes\Pi _b)U^\dag\right),$$ which
acts on the quantum system $S$ alone. This operator is
positive-semi\-definite, and a family of such operators form
resolution of identity. They are not, however, mutually
orthogonal. Such operators form positive operator-valued measures
(POVM)\cite{Peres}.

\section{Conclusion}

We have shown how to obtain a description of quantum measurement
via POVM at the condition of disregarding completely the physical
interaction during measurement and the existence of P-observer. If
one is only interested in a formal description of how I-observer
acquires information about the quantum system $S$, this is done
via POVM and the Born rule following from Gleason's theorem. To be
mentioned here, Gleason's theorem also admits a generalization
from von Neumann's orthogonal projector measures to POVM
\cite{Busch}. One gets therefore a description of measurement as
used in quantum information theory, and one can now continue the
development of the theory in the conventional way\cite{Peres}. In agreement with the intuition expressed in the
key metaphor, all "hardware" language is eliminated and the theory
can be formulated in the "software" language alone.

Formal deduction of the results concerning the Hilbert space,
however, was not completely rigorous. Rovelli\cite{RovRQM}
acknowledges it in his disclaimer, "I do not claim any
mathematical nor philosophical rigor". Indeed, the fact that
yes-no questions form an orthomodular lattice containing subsets
that form Boolean algebras only commits one to the structure of
union of Hilbert spaces and not of a single Hilbert space. Thus,
this can happen to be the union of primitive Hilbert spaces, which
allow for a classical and not a quantum interpretation. Generally
speaking, the structure will be the one of the Hilbert space with
superselection rules. One needs then to use Axiom 2 to show that
the possibility to ask in every situation some new informative
question excludes classicality. Completion of this program remains
an open problem.

Introduction of space and time "by hand" in any algebraic approach
to quantum mechanics is certainly quite unsatisfactory. One would
wish to see how space and time arise naturally from the formalism.
This may require a fully rigorous algebraic approach involving von
Neumann algebras and the GNS construction for the Hilbert space.
However, the author is only aware of one way to introduce time in
this framework\cite{ConnesRovelli}. This leaves open the question
of link between time and space, and of the possibility to use the
two notions together to obtain the evolution equation.

\section*{Acknowledgements}

The author would like to thank Carlo Rovelli and Christopher Fuchs
for discussion and suggestions.

\end{document}